# Self pinning protein-laden drops

Viatcheslav V. Berejnov

Physics Department, Cornell University, Ithaca, NY, 14853-2501

berejnov@uvic.ca

Present address: IESVic, University of Victoria, PO Box 3055, Victoria, BC V8W 3P6, 250-472-4202

ABSTRACT

**Proteins dissolved in a drop induce and enhance the pinning of the drop contact line. This effect dramatically increases volume of drops that are vertically pinned on a flat siliconized substrate. The drop pinning behavior exhibits two regimes: for low protein content in a drop the pinning increases as the contact angle hysteresis increases, and for high protein content the pinning decreases as the surface tension of the protein solution decreases.**

## Introduction

Surface-active molecules in solution have the potential to change the material and mechanical equilibriums in the area of a gas-liquid-solid contact line. Particularly, by changing either the chemical nature of the surfactant or concentration, it is possible to modify wetting, spreading, and contact angle hysteresis of a liquid. The effect of surfactant concentration on droplet wetting behavior has been well studied experimentally [1-14]. For flat solid surfaces it was observed that these surfactants may extend the spreading rate [8-10, 12] by controlling i) the stability of the advancing contact line, the shape of which may span from continuous to dendritic [13, 14] and ii) the motion of the contact line that may exhibit the monotonous or "stick-jump" type of invasion along a solid surface [14]. It was found that low molecular weight surfactants may transport toward and through the contact line that leads to self-assembly and rearrangement of the surfactant [2-4, 6]. The last may induce the so-called "autophobic" effects when the surfactants modify the property of the gas-solid surface to provide spontaneous dewetting of a contact line [6]. For super-hydrophobic surfaces it was found that the adsorption of surfactants from the bulk may also control the mode of wetting ranging between Wentzel and Cassie regimes [5]. Recently, the rigorous modeling of surfactant solution spreading over a hydrophobic flat surface was attempted [7]. Their model accounted for the transport of the surfactant through the contact line, and reasonable agreement between theory and experiment was demonstrated [7].

In this article we report that proteins (bio-macromolecular surfactants) interact differently with a siliconized hydrophobic surface compared to low molecular weight surfactants. We found that proteins stabilize and pin the three-phase contact line with high efficiency. This effect of dissolved proteins in a drop on contact line pinning has not yet been thoroughly discussed in the literature despite being a primary issue for several important methods in life sciences. The optimization of drop pinning conditions could benefit the crystallization of globular and membrane proteins [15, 16], the formulation of pesticide sprays for protecting the plants [1, 17], and the influence of pulmonary surfactants on the physiological wettability of alveoli in lungs [18], particularly since all these methods deal with curved liquid-fluid meniscuses adhered to a solid substrate.

In this article, we study quasi-static pinning of protein drops on siliconized glass slides. The effect of pinning was measured in terms of volumes of drops statically anchored by vertically tilted slides. These drops were considered to be completely pinned. Proteins were dissolved in the drops and affected the wetted part of slides in contact with the drop's interior. We found that the drop pinning behaviour differs between low and high protein concentration regions, respectively. For low concentrations, proteins dissolved in drops increase the contact angle hysteresis and enhance the volume of vertically pinned drops by a factor of 4-

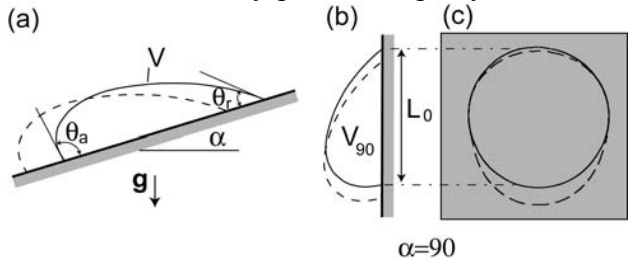

**Figure 1** Schematic of the inclined and vertical drops and notations. Onset of pinning(depinning) is depicted by the solid and dashed drop profiles, which correspond to constantly pinned and continuously moving drops, respectively. (a) is a geometrical sketch of an apparatus for measuring the stability diagrams for inclined pinned drops. (b) and (c) are side and top views of a vertically pinned(depinned) drop, respectively. The typical deformation of the contact line of the depinned and continuously sliding drop is shown on (c). The contact line of a pinned drop at equilibrium is approximated by a circumference with a diameter $L_0$.

5 (compared to water or a buffer solution). In the high concentration region, when the contact angle hysteresis seems to be saturated, the surface tension controls pinning by decreasing the volume of pinned drops with respect to the protein concentration.

**Experimental and Methods**

In our experiments, DI water was purified by NANOpure II (Barnstead, Boston, MA). Lysozyme protein (Lys), 6-x times crystallized hen egg white, was purchased from Seikagaku America (Mr~14 kD, lot: LF1121, Falmouth, MA). Lys was dissolved in a 50 mM sodium acetate buffer with pH = 5. All solutions were filtered through 0.45 μ Millipore filters. The protein concentration was measured by measuring the mass of the components. For some samples we used a Spectronic Genesys 5 (Waltham, Ma) spectrophotometer at a wavelength of 278 nm to measure the actual protein concentration in solution through light absorption; the maximum difference between the mass and light absorbance methods was ~ 5%. The extinction coefficient used for Lys in the optical method was 2.64 mL·(mg·cm)$^{-1}$. The second virial coefficient $A_2$ for Lys molecules in our buffer was ~ $10·10^{-4}$ (mL·mol/g$^2$) [19]. This positive value of the second virial coefficient demonstrated a high degree of repulsion between proteins in the buffer medium [20, 21], which decreased their self assembling rate on the solution-air interface and stabilized the solution surface tension.

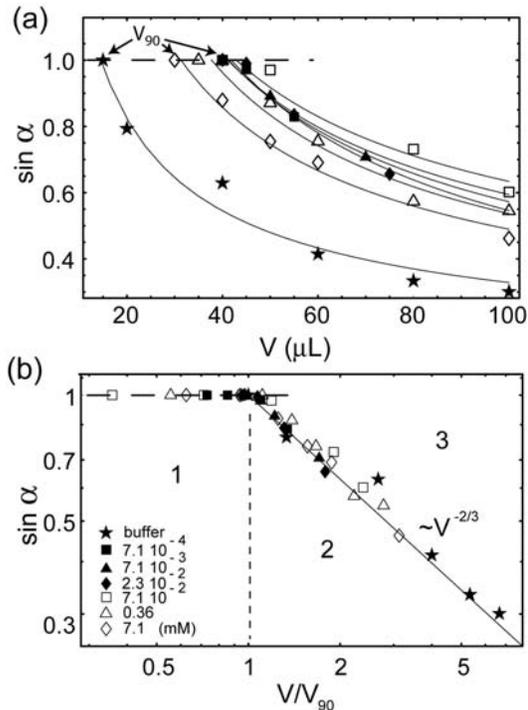

**Figure 2** Stability diagrams of drops pinned on the siliconized hydrophobic glass slides for different protein solutions, (a), and their log-log representations, (b). On (a) for given Lys concentration the drop is stably pinned if volume, V, and inclination, α, are less than critical provided by the correspondent curve; every curve provides a maximum volume, $V_{90}$, of a drop pinned by the vertically inclined slide. The dashed line denotes an inclination at α=90°. On (b) the V-axis is normalized by $V_{90}$ and data is collapsed to the fitting line ~ $V^{-2/3}$; the numbers denote the zones where (1) the drop is absolutely stable against any inclinations, (2) the drop is stable up to inclination <90°, and (3) the drop is unstable and moves continuously.

The surface tension γ for the Lys solutions was measured using a pendant drop counting method [15], with an experimental accuracy of ~ 5-7%. We used plastic hydrophobic tips to dispense drops. The time interval between drops was $\tau_\gamma$ ~ 1 min. Spontaneous equilibration [22-24] of the newborn interface of protein solutions affects both the γ–measurements and the experiments of dispensing/inclining the protein drops. This equilibration leads to the appearance of an induction time, which is the time of characteristic evolution of surface tension of a new interface. Experiments with Lys dissolved in buffers similar to ours with a positive virial coefficient demonstrated a low protein assembling rate and consequently a small decay of surface tension of the newborn interface within the first hour of its existence. We estimate a low bound of induction time in $\tau_i$ > 60 min [22, 23]. Our pinning experiments with inclined drops did not exceed 20 min ($\tau_d$ < 20 min), which roughly corresponds to decreasing the surface tension due to spontaneous equilibration on ~ 1 mN/m for the Lys concentration of ~ 7.1 $10^{-2}$ mM (1 mg/mL) [22]. Therefore, we can conclude that the measurements of surface tension report the true values in the period of measuring the critical volumes of the inclined drops.

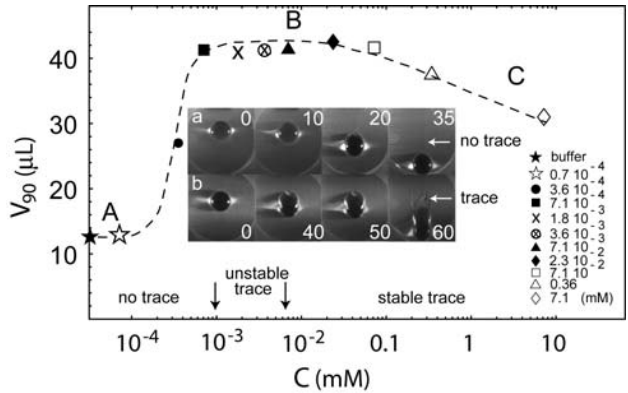

**Figure 3** Stability diagram of the vertically pinned drops. Dashed curve is an eye-guide. The zones above and below the curve correspond to continuously moving and stably pinned drops, respectively. The inset shows the appearance of a trace behind the drop body with respect to an increase of Lys content in the drop; numbers denote an inclination in degrees, drop volume is 50 μL; a – buffer solution, b – protein solution, 0.36 mM of Lys. The letters A, B, and C depict the three characteristic regimes of pinning with respect to the Lys concentration, see text.

Flat, siliconized 22 mm glass slides HR3-231 were purchased from Hampton Research (Laguna Niguel, CA). A water drop with a volume of ~20 μL dispensed on a new slide formed a reproducible advancing contact angle of ~ (92±1)°. The details of the water drop contact angle hysteresis was presented elsewhere [25]. According to the Hampton Research Customer Support, the siliconized material of the HR3-231 slides was similar to the organosilane-composed solution AquaSil (Hampton Research). We inspected the surface topography for some of the HR3-231 slides using a contact mode AFM (DI MultiMode III, Santa Barbara, CA) with NSC 1215 tips from MikroMasch. We found the manufacturer's coating to be homogeneous and flat.

Each drop was manually dispensed onto an initially horizontal glass slide. A goniometer table was slowly rotated, Figure 1a, until the pinned drop approached a critical zone of inclination [25]. In this zone the tilt increased in ~ 2° steps. The relaxation time between rotations in the critical zone was ~ 30 s to allow the transient mechanical disturbances of the drop to dissipate. The apparent advancing, $\theta_a$, and receding, $\theta_r$, contact angles, Figure 1a, were measured for some drops at quasi-static equilibrium using the same goniometer. Dispensed drop volumes were accurate to 0.1 - 0.5 %, and tilt and contact angle measurements were accurate to 1 - 2°.

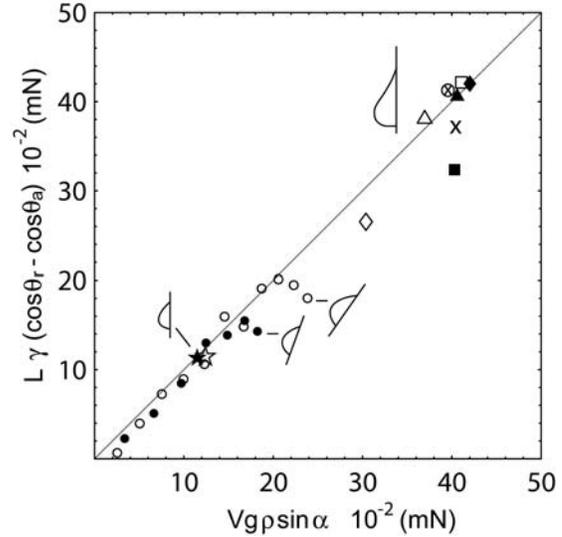

**Figure 4** Comparison of the Lys solutions to Eq. (1). The solid line is a fit to Eq. (1). Symbols ● and ○ depict pinning of 20 μL and 30 μL water drop at their progressive inclination [25]; the last points in these series correspond to the last stably pinned drops at inclinations shown on the inserts, respectively. Other symbols correspond to data presented in Figures 2 and 3; the two inserts with the vertically pinned drops demonstrate typical drop profiles observed in the A (★, ☆), B (x, ⊗, ▲) and C (◆, □, △, ◊) zones, respectively.

We did not observe the autophobic [14] and Marangoni-induced [13] contact line displacements and the decreasing advancing contact angles at high protein concentrations [26, 27].

The drop pinning was characterized by measuring the critical tilt $\alpha$ corresponding to the onset of continuous motion of an inclined drop of volume V [25], Figure 1. The drop stability diagram (V, $\alpha$) was a "fingerprint" of the pinning conditions corresponding to a particular substrate and protein solution. Thus, we have collected these (V, $\alpha$) diagrams for different Lys concentrations C; see Figure 2 (a). We observed that the scaling law $\sin(\alpha) \sim V^{-2/3}$, which can be deduced from [28], reasonably fits the (V, $\alpha$) data over a broad range of protein concentrations, Figure 2 (b). The curves in Figure 2 (a) separate a region of completely unstable drops from the region, where drops either were absolutely stable or eventually were getting a stable configuration of the contact lines after series of transient displacements [25]. Volumes $V_{90}$ indicated the largest drops of different protein concentrations, which were statically pinned by the siliconized slide tilted at 90°, Figures 1b, 1c, and 2 (a). We found parameter $V_{90}$ to be very convenient, which is the fitting product of the (V, α) diagram, because it robustly described the effect of proteins on pinning the drop contact line, Figures 2 and 3. A function (C, $V_{90}$) was taken to characterize pinning of the vertically inclined drop with respect to the content of proteins dissolved in the drop aliquot, Figure 3.

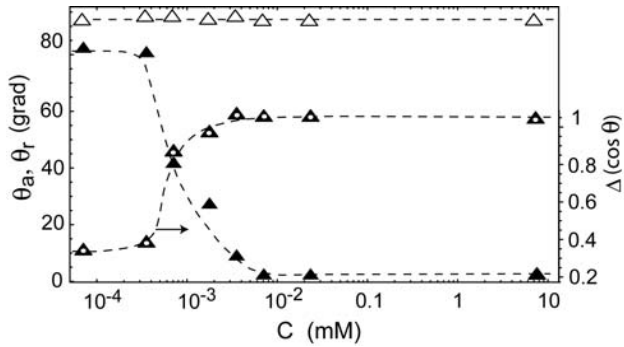

**Figure 5** The maximal and minimal contact angles (left) and the contact angle hysteresis (right) of protein solutions with respect to the Lys concentration. The contact angles correspond to the static drops at critical inclinations. Dashed lines are the eye-guides. Symbols denote △: advancing angles, $\theta_a$; ▲: receding angles, $\theta_r$; ◬: contact angle hysteresis, $\cos\theta_r - \cos\theta_a$.

**Results and Discussion**

The curves presented in Figure 2 (a) map the zones of drop equilibrium and constant motion down to an inclined substrate for different protein contents in the drops. At its equilibrium, the drop contact line was not displaced at all, or when it was displaced it was always able to find a new stably pinned configuration. We term this behavior of the drop as the stably pinned drop and we do not discuss in this article the transitional pinning regimes considered elsewhere [25]. On Figure 2 (a) the zones below each equilibrium curve correspond to the stably pinned drop, zones 1 and 2 on Figure 2 (b), whereas the zones above correspond to the continuously moving drop (stably depinned drop), zone 3 on Figure 2 (b). By interpolating the equilibrium curves to the vertical tilt α=90°, we found maximum volumes $V_{90}$ of vertical stably pinned drops for different protein concentrations, C, in the drops. The $V_{90}$ values correspond to drops at equilibrium, thus we may treat the plot presented in Figure 3 as a stability map of pinning, which shows the stability of a vertically pinned drop with respect to the protein content in the drop. Similarly to the plots in Figure 2 the zones below and above the dashed curve in Figure 3 correspond to the stable drop pinning and stable drop depinning zones, respectively.

Several different regimes of pinning with respect to the protein concentration can be seen in Figure 3. For very low concentrations < $10^{-4}$ mM, an interval A, pinning of the Lys-laden drops behaves similarly to the pure buffer solution or water drops [25]. For these concentrations the contact line depins without leaving a noticeable liquid trace behind the drop body. Increasing the concentration of Lys in solution up to ~ $10^{-3}$ mM yielded a dramatic increase of volumes of vertically pinned drops, which we interpreted as an increase of pinning. In the large concentration interval, zone B (between ~ $10^{-3}$ mM and ~ 0.1 mM), the pinning seemed to be saturated. We observed an appearance of liquid traces behind the drops in this interval. These traces were unstable in low Lys concentrations; they grew with respect to the drop displacement and abruptly lost their pinning stability simultaneously over the entire circumference of the receding part of the contact line, which provoked the onset of continuous motion. For Lys concentration ~ $10^{-2}$ mM, the stability of the traces was noticeably increased making them a stably pinned film-like pattern behind the drop body. A further increase of Lys concentration > 0.1 mM (zone C) resulted in the slow decrease of pinning of vertical drops. No morphological changes of the drop profile including the trace part were observed after ~ 0.1 mM of Lys.

Equilibrium of an inclined drop yields an equality of two "net" forces projected on the surface of the inclined substrate [29, 30]:

$$\rho V g \sin(\alpha) = L\gamma\Delta(\cos\theta) \qquad \text{Eq. (1)}$$

where ρ, V, g, γ, L, and $\Delta(\cos\theta)=\cos\theta_r-\cos\theta_a$ are the density of liquid, volume of the drop, gravity, surface tension, the characteristic drop diameter, and the contact angle hysteresis, respectively. Vertically inclined drops are a special case described by Eq. (1) and, thus, this equation may be applied to the equilibrium map presented in Figure 3. To complete the data required by Eq. (1) the surface tensions and contact angles for experiments presented in Figure 3 were collected. The complete data set corresponding to Figure 3 is presented in Figure 4 within the axes natural for Eq. (1). The solid line in Figure 4 shows an ideal stability of an inclined drop, see references elsewhere for details [1, 25, 29, 30]. Rounded symbols denote the stability of the water drops of 20 μL (closed) and 30 μL (opened) taken from [25] for comparison with the Lys data corresponded to the vertically inclined drops.

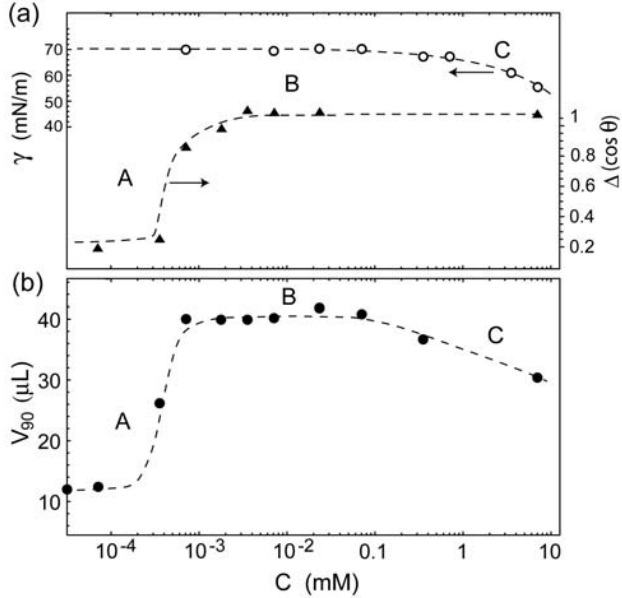

**Figure 6** Combination of plots of the contact angle hysteresis and surface tension, (a), with the diagram of stability, (b), *vs.* protein concentration. The frame (a) represents the surface tension data: ○ that is at left and the contact angle hysteresis $\Delta(\cos\theta)$: ▲ that is at right *vs.* the Lys concentration. The frame (b) represents the volume of vertically pinned drops, $V_{90}$: ● with respect to the Lys concentration (Figure 3). Dashed lines are the eye-guides. The letters A, B, and C depict three characteristic regimes of pinning with respect to the Lys concentration presented in Figure 3.

Figure 4 demonstrates several issues discussed below. First, the simple approach proposed in Eq. (1) may be used as a measure of the pinning effect only if the inclination angles of the compared inclined drops are fixed. Otherwise, the progressive inclination changes the pinning conditions, making comparison difficult. Second, the deviation of data from the solid line may reflect the importance of transitional displacements that the contact line undergoes in order to reach a more stable configuration. These displacements change the characteristic drop diameter, L. For more details, see [25]. The data corresponding to water drops in Figure 4 support these two conclusions. Thus, we account for the deviation between the solid line and the Lys solution points by the changing characteristic drop diameter, L. Third, the data points for the vertically pinned drops of monotonically increasing protein concentration are distributed along the solid line in a non-monotonic trend. The critical points of highest protein concentrations are between the points corresponding to the lowest and moderate concentrations. For this region two protein concentrations may provide the same point on the solid line. For two drops with the same volume, according to Eq. (1), this coincidence is possible if parameters: $\gamma$ and $\Delta(\cos\theta)$ yield the same product $\gamma\Delta(\cos\theta)$ for varying protein concentrations.

By generalizing the concentration dependence we can study the following two correlations:

$$V_{90} \sim \cos\theta_r(C) - \cos\theta_a(C) \quad \text{and}$$
$$V_{90} \sim \gamma(C), \qquad \text{Eq. (2)}$$

where we assume that the surface tension and contact angles depend on protein concentration but are independent from each other. Other parameters included in Eq. (1) did not depend on protein concentration. The characteristic drop diameter, L, (which was close to an unperturbed drop diameter $L_0$ in our experiments, Figure 1) is not a material property and does not directly depend on concentration, but the function of the initial conditions of the drop either dispensing or displacing.

Figure 5 and 6 show the correlation data introduced by Eq. (2). We can see that Lys proteins affect both the contact angles and surface tension but in different concentration intervals. Particularly, at the low protein concentration, starting at $10^{-4}$ mM, the contact hysteresis increases as concentration increases, reaching its maximum and saturates, Figure 5. The surface tension begins to be affected by the proteins in a concentration range that is three orders of magnitude higher, between 0.1-1.0 mM, Figure 6 (open circles). The fact that the contact angles and surface tension are affected by proteins in different concentration intervals supports the claim that the introduced above functions of $\theta(C)$ and $\gamma(C)$ are independent one from another.

Applying the contact angle hysteresis and surface tension correlations to the pinning data we can see a remarkable coincidence, Figure 6. The pinning of the vertically inclined drops increases as the contact angle hysteresis increases, reaches its maximal value as the hysteresis saturates, and decreases as the surface tension decreases. Figure 5 shows that proteins affect the advancing and receding parts of the drop contact line differently. The increase of pinning corresponds to the increase of contact angle hysteresis, which is attributed to the decrease in receding contact angle with respect to the drop protein concentration. It is interesting to note that in this concentration interval, we observed an appearance of liquid traces behind the drop.

Two effects of protein concentration on the contact line presented on Figure 5 need to be understood: an abrupt decrease of the receding angle, $\theta_r$, and an independence of the advancing angle, $\theta_a$.

The presence of liquid film-like traces behind the drop during its quasi-static displacement indicates a high degree of anchoring of the receding contact line to the solid surface (see the insert on Figure 3). This anchoring depends on the protein concentration in the drop deposed on the substrate (see the $\theta_r$ on Figure 5). It is reasonable to assume an adsorption of proteins to the substrate [23, 26, 31, 32]. We verified this adsorption by first rinsing the area of the substrate wetted by the protein solution with DI water, SDS solution, and NaCl solution, sequentially. The DI water drops dispensed on this rinsed area indicated a higher hysteresis compared to the native substrate area. Thus, we can conclude that some lysozyme irreversibly adheres to the substrate. While Lys concentration reached ~ 7.1 $10^{-3}$ mM in the bulk, the substrate surface acquired enough adsorbed proteins to anchor the receding contact line, decreasing the receding contact angle, $\theta_r$, to its minimum.

Independence of the static advancing angle, $\theta_a$, in Figure 5 on the Lys concentration indicates the low (or zero) transport of proteins in the region of close proximity to the contact line. While the contact line of the protein laden-drop displaces in the advancing direction, the contact angle is the result of the interaction of the buffer component and the native siliconized surface (see the $\theta_a$ at ~ 0 mM on Figure 5). Two reasons may support this hypothesis. First, the fastest mode of adsorption of proteins to solid is dominated by diffusion [31], which takes place between ~ 1-10 sec, while the drop displacement time is faster ~ milliseconds in our case. Thus, during the time that the contact line displaces, the buffer comes into contact with the solid first. Second, we need to assume that proteins cannot equilibrate the drop's gas-liquid and liquid-solid interfaces in close proximity to the contact line edge, while they can equilibrate those interfaces far away from the contact line. We speculate that the proteins deplete each other from the contact line region, since they have a tendency to be oriented differently in close proximity to the gas-liquid or liquid-solid interfaces. The fact that the surface of the protein globule is covered by the domains of different chemical affinity [16], which interact differently with the gas-liquid and liquid-solid interfaces, may support this hypothesis. Few to no molecules match the energetic conditions of the gas-liquid-solid contact edge. Therefore, the depleted zone filled with buffer only appears in close proximity to the advancing contact line, defined by $\theta_a$. This explanation is reasonable to the nature of proteins, buffers, substrate surfaces.

**Conclusion**

Summarized below are our results that correspond to the effect of enhancement of contact line pinning from dissolved proteins. Dissolved proteins dramatically increase drop contact line pinning on a siliconized glass substrate. By mapping the stability curves of the inclined drops, we characterized pinning for different protein concentrations in the drops. The function of volume of the vertically inclined drops with respect to the protein concentration was used for measuring pinning conditions. It was shown that pinning behaves non-monotonically depending on the concentration interval. Two functions, the contact angle hysteresis and the surface tension versus concentration, were measured to study the causes of the pinning effect. The remarkable correlation between these functions and the drop pinning is established in the appropriate concentration intervals. It was shown that contact angle hysteresis and surface tension affect pinning in noticeably different zones of protein concentration.